%% file: ACCESS2017.tex
\documentclass[journal]{IEEEtran}
\usepackage{balance}
\usepackage{etex}
\input{text/support}

\begin{document}
	\input{text/title}
	\input{text/abstract}
	\input{text/introduction}
	\input{text/relatedwork}
	\input{text/model}
	\input{text/npcomplete}
	\input{text/algorithms}
	\input{text/experiments}
	\input{text/conclusion}

	\bibliographystyle{unsrt}
	\bibliography{social,pids,myref,triangle,vul,cnd,myref2,mendeley}
	\balance
	\newpage
	
	\input{text/expfigures}

\end{document}

%% file: text/title.tex
\title{Transitivity Demolition and the Falls of\\ Social Networks}

\author{Hung~T.~Nguyen,~Nam~P.~Nguyen,~Tam~Vu,~Huan~X.~Hoang,
        and~Thang~N.~Dinh,~\IEEEmembership{Member,~IEEE}
\thanks{Hung T. Nguyen and Thang N. Dinh are with the Computer Science Department, Virginia Commonwealth University, Richmond, VA, 23220 Email: \{hungnt, tndinh\}@vcu.edu.}
\thanks{Nam P. Nguyen is with Computer and Information Sciences Department, Towson University, Towson, MD, 21252 Email: npnguyen@towson.edu.}
\thanks{Tam Vu is with Computer Science and Engineering Department, University of Colorado, Denver, CO, 80204 Email: tam.vu@ucdenver.edu.}%
\thanks{Huan X. Hoang is with Information Technology Department, Vietnam National University, Hanoi, Vietnam Email: huanhx@vnu.edu.vn.}%
\thanks{Manuscript received ; revised ; accepted.}}



\maketitle

%% file: text/abstract.tex
\begin{abstract}
In this paper, we study crucial elements of a complex network, namely its nodes and connections, which play a key role in maintaining the network's structure and function under unexpected structural perturbations of nodes and edges removal.
Specifically, we want to identify vital nodes and edges whose failure (either random or intentional) will break \textit{the most number of connected triples (or triangles)} in the network.
This problem is extremely important because connected triples form the foundation of strong connections in many real-world systems, such as mutual relationships in social networks, reliable data transmission in communication networks, and stable routing strategies in mobile networks.
Disconnected triples, analog to broken mutual connections, can greatly affect the network's structure and disrupt its normal function, which can further lead to the corruption of the entire system.
The analysis of such crucial elements will shed light on key factors behind the resilience and robustness of many complex systems in practice.		
	
We formulate the analysis under multiple optimization problems and show their intractability. We next propose efficient approximation algorithms, namely \IN{} and \IE{}, which guarantee an $(1-1/e)$-approximate ratio (compared to the overall optimal solutions) while having the same time complexity as the best triangle counting and listing algorithm on power-law networks.
This advantage makes our algorithms scale extremely well even for very large networks.
In an application perspective, we perform comprehensive experiments on real social traces with millions of nodes and billions of edges.
These empirical experiments indicate that our approaches achieve comparably better results while are up to 100x faster than current state-of-the-art methods.
\end{abstract}


\IEEEpeerreviewmaketitle

%% file: text/introduction.tex
\section{Introduction}
\IEEEPARstart{R}{obustness} and resilience to unexpected perturbations is perhaps one of the most desirable properties for corporeal complex systems, such as the World Wide Web, transportation networks, communication networks, biological networks and social information networks. 
In general, resilience of a network evaluates how much the network's normal function is affected in case of external perturbation, i.e., it measures the network in response to unexpected events such as adversarial attacks and random failures \cite{holme2002attack}.
In order to improve the robustness of real-world systems, it is therefore important to obtain key insights into the structural vulnerabilities of the networks representing them.
A major aspect of this is to analyze and understand the effect of failure (either intentionally or at random) of individual components on the degree of clustering in the network.

Clustering, or more particularly, \textit{the number of connected triples/triangles}, is a fundamental network property that has been shown to be relevant to a variety of topics, such as communities of genes in biological networks, forwarding and routing tables mobile networks, and especially strong connection of users in online social networks (OSNs) \cite{watts1998cds}.
Connected triples nicely capture the social intuition ``a friend of your friend is also your friend'' \cite{Leskovec:2007:DVM:1232722.1232727}, and thus, is the fundamental pattern of information diffusion in multiple systems.
For example, consider the propagation of information through a social network, such as the spread of a rumor.
A growing body of work has identified the importance of the number of connected triples to such propagation; the more connected triples a network has, the easier it is for information to propagate \cite{centola2010spread, barclay2013peer,lu2011small,malik2013role,centola2011experimental}.
Connected triples are also behind the fall of some online social sites, such as MySpace and Friendster, as they suffered a catastrophic degrade of active users, activity traffic, and consequently, popularity in the cyberspace. For instance, Friendster claimed to have over 100 million users at its peak, but most them had quit and fled to other networks (e.g., Facebook) by the end of 2009 \cite{msft01, cyberlaw}, triggering a cascade of broken bonds and friends leaving Friendster.
The identification of elements that crucially affect the number of connected triples in the network, as a result, is of great impact.

The importance of connected triples is not limited to social networks; in the context of air transportation networks, \cite{Ponton2013} argued that those connected triples of such a network is beneficial, as passengers for a canceled flight can be rerouted more easily.
This metric also plays an important role in the network community structure, which is the core of mobile forwarding and routing strategies in Delay Tolerant Networks (DTNs). Particularly, \cite{NamTMC2016} has shown the correlation between the number of disconnected triples and the significant degrade of forwarded packets in DTNs.
In addition, as a matter of homeland security, the critical elements for clustering in homeland communication networks should receive greater resources for protection; in complement, the identification of critical elements in a social network of adversaries could potentially limit the spread of information in such a network.

Many measures have been proposed for evaluating the resilience of technological and biological systems; however, there are only few work suggested for social networks.
Most studies in the literature focus on how the network behaves under perturbation using the measures such as the pair-wise connectivity \cite{thangton2012}, natural connectivity \cite{hau2014}, or using centrality measures, e.g., degree, betweeness \cite{Albert00theinternet}, the geodesic length \cite{holme2002attack}, eigenvector \cite{allesina2009}, etc.
Nevertheless, most of them (1) focus only on the local but not the global network' structure, and (2) do not take mutual interactions and social relationships into account. These limits drive the need for another metric for social resilience.
To our knowledge, none of the existing work has examined the number of connected triples from the perspective of vulnerability - as evidenced by the examples above, the damage made by the broken triples, resulted from element-wise failures, can potentially have severe effects on the functionality of the network.
This drives the need for an analysis of this metric in complex networks.

Our study in this paper investigates the structural resilience of OSNs under the scenarios of element-wise failures, particularly under two scenarios of adversary attacks and random failures. Our goal is to discover and protect critical network' elements (nodes and links) whose failures will break most triples in the network. In a nutshell, our contributions are
\begin{enumerate}
\item We study the resilience of social networks through the number of connected triples. This an important structural vulnerability of an OSN that can greatly affect its popularity among the crowds. We formulate the analysis under multiple optimization problems, and show their hardness and intractability.
\item We propose efficient approximation algorithms to identify triangle-breaking points (i.e., nodes and links) in the network structure: \IN{} algorithm for node removal and \IE{} algorithm for edge removal. Our proposed approaches guarantee are a small constant factor in comparision to optimal solutions. Interestingly, both \IN{} and \IE{} have the same time complexity with the best triangle counting/listing algorithms, $O(m^{\frac{3}{2}})$. This makes our algorithms scale extremely well for large social data.
\item We also investigate the \textit{input-dependent bounding} technique previously appeared in \cite{Leskovec07} for influence maximization problem. The input-dependent bound usually gives better approximation guarantee than the worst-case bound since it accounts for the particular instance of the problem and particular run of the algorithms. As shown in the experiments, the input-dependent bound vastly improves over the worst-case guarantee and for some networks, returns the exact optimal solutions.	
\item We carry out extensive experiments in comparison with state-of-the-art methods on real-world data with millions of nodes and edges. The results show that \IN{} and \IE{} substantially outperform the other algorithms in terms of running time: up to 100x faster than the direct competitor, GreedyAll \cite{Li14}, which was shown to have the best solution quality and be among the most scalable methods in their papers.
\end{enumerate}    
Paper organization: Section \ref{ssrelatedwork} reviews studies that are related to our work. Section \ref{sspreliminary} describes the notations, measure functions and problem definitions. Sections \ref{sshardness} shows the proof of NP-completeness implying the intractability of these investigating problems. Sections \ref{sec:algs_node} and \ref{sec:algs_edge} present our solutions \IN{} and \IE{} for the problems of interested, respectively. In section \ref{ssexperiments}, we report empirical results of our approaches in comparison with other strategies. Finally, section \ref{ssconclusion} concludes the paper.

\begin{table*}[t]
\centering
\caption{List of Symbols}
\begin{tabular}{ll}
	\addlinespace
	\toprule
	Notation  &  Meaning \\
	\midrule
	$n$ & Number of vertices/nodes ($N = |V|$)\\
	$m$ & Number of edges/links ($M = |E|$)\\
	$d_u$ & The degree of $u$\\
	$N(u)$ & The set of $u$'s neighbors\\    
	$Tri(u)$ & The set of triangles on a node $u$\\    
	$T(u)=|Tri(u)|$ & The number of triangles on $u$\\
	$Tri(u, v)$ & The set of triangles on an edge $(u, v)$\\    
	$Tri(S) = \displaystyle\cup_{u \in S} Tri(u)$ & The set of triangles on $S \subseteq V$\\
	$Tri(F) =  \displaystyle\cup_{(u, v) \in F} Tri(u, v)$ & The set of triangles on a subset of edges $F \subseteq E$\\            
	\bottomrule
	\end{tabular}
	\label{tab:syms}
\end{table*}

%% file: text/relatedwork.tex
\section{Related Work}
\label{ssrelatedwork}
Many metrics and approaches have been proposed to account for network robustness and vulnerability \cite{Grubesic08,  Murray08,  Neumayer11, Dinh11, Dinh15}. While each of these measures has its own emphasis and rationality, they often come with several shortcomings that prevent them from capturing desired characteristics of network connectivity and resilience. For example, measures based on shortest path are rather sensitive to small changes (e.g. removing edges or nodes); algebraic connectivity and diameter are not meaningful for disconnected graphs (all disconnected graphs have the same values); number of connected components and component sizes, arguably, do not fully reflect level of network connectivity.
		

Vulnerability assessment has attracted a large amount of attention from the network science community. Work in the literature can be divided into two categories: Measuring the robustness and Manipulating the robustness of a network. In measuring the robustness, different measures and metrics have been proposed such as the graph connectivity \cite{thangton2012}, the diameter, relative size of largest components, and average size of the isolated cluster \cite{Albert00theinternet}. Other work suggests using the minimum node/edge cut \cite{Frank1970} or the second smallest non-zero eigenvalue or the Laplacian matrix \cite{Fiedler73}. In terms of manipulating the robustness, different strategies has been proposed such as \cite{Albert00theinternet}\cite{peixoto2012}, or using graph percolation \cite{Callaway2000}. Other studies focus on excluding nodes by centrality measures, such as betweeness and the geodesic length \cite{holme2002attack}, eigenvector \cite{allesina2009}, the shortest path between node pairs \cite{Grubesic08}, the pair-wise connectivity \cite{thangton2012}, propagation of worms and cascading failures \cite{Nguyen2010novel, Dinh2012cheap, Dinh2014cost}. More information of general vulnerability assessment can be found in \cite{hau2014} and references therein.
	
Community structure \cite{Nguyen11socialcom} is an another common pattern found in real-world networks. Network structural vulnerability in social networks, has so far been an untrodden area.  In a related work \cite{namASONAM13}, the authors introduced the community structure vulnerability to analyze how the communities are affected when top $k$ vertices are excluded from the underlying graphs. They further provided different heuristic approaches to find those critical components in modularity-based community structure. \cite{namWI14} suggested a method based on the generating edges of a community to find the critical components.
	
Counting and listing triangles in a graph is an important problem, motivated by applications in a variety of areas. The problem of counting triangles on a graph with $n$ vertices and $m$ edges can be performed in a straightforward manner in $O(mn)$. This has been improved to $O(m^{3/2})$ in \cite{Schank05} and $O(m^{\frac{2w}{w+1}})$ where $w < 2.376$ is the exponent of matrix multiplication \cite{Alon97}. To improve the performance of triangle counting in large graphs, parallel algorithms are also studied in \cite{Suri11}. There are also several works on approximate triangle counting \cite{Bar02,Buriol06,Jowhari05}. Recently, the \kN{} and \kE{} problems are investigated in \cite{Li14}. The authors provides NP-completeness proofs and greedy algorithms for the problems. Unfortunately, the NP-completeness proofs contains fundamental flaws that cannot be easily fixed.

%% file: text/model.tex
\section{Model and Problem Definition}
\label{sspreliminary}
In this section, we first describe the main problem of interest, and then define its four triangle-breaking variants. We then prove the NP-hardness of those problems. Based on the submodularity property of the objective functions, the approximability is stated accordingly for each problem based on the rich literature of optimizing submodular functions \cite{Gandhi04,Ageev04}.
	
We represent a social network by an undirected graph $\mathcal G=(V, E)$ with $|V|=n$ nodes and $|E|=m$ undirected edges. Given a graph $G=(V, E)$, we study multiple attack models in which the attackers attempt to break the most number of triangles in the graph by removing nodes and edges either intentionally or at random. Here, a triangle is broken if one of its edges or nodes is removed from the graph. In what following, we define four variants of the triangle-breaking problem based on Node and Edge removals. 
    
\subsection{Problem Definition}	
	
\begin{definition}[$k$-triangle-breaking-node]
Given an undirected graph $G=(V, E)$ and budget size $k$, find a subset $S^*$ of $k$ nodes whose removal will break the maximum number of triangles in $G$
\begin{align}
	S^* = \argmax &\quad \quad |Tri(S)| \nonumber\\
	\text{s.t.} &\quad \quad |S| \leq k, \nonumber\\
	&\quad \quad S \subseteq V \nonumber
\end{align}
where $Tri(S)$ is the set of triangles with at least one node in $S$, i.e., 
\begin{align}
	Tri(S) = \{ (u, v, w) \ &|\ (u, v), (v, w), (w, u) \in E \nonumber\\
	&\text{ and } \{u, v, w \} \cap S \neq \emptyset \nonumber\}.
\end{align}
\end{definition}
	
Note that we can formulate the above problem as an Integer Linear Programming problem (ILP). For each $u \in V$, define $x_u \in \{0, 1\}$ such that
\begin{align}
	x_u = \left\{ \begin{array}{ll}
		1 & \text{ if node } u \text{ is removed,}\\
		0 & \text{ otherwise}.
		\end{array}
		\right.\nonumber
\end{align}
and for each triangle $(u, v, w) \in Tri(V)$, define an integral variable $y_{uvw} \in \{ 0, 1\}$ that satisfies
	\begin{align}
		y_{uvw} = \left\{ \begin{array}{ll}
			1 & \text{ if triangle } (u, v, w)  \text{ is broken,}\\
			0 & \text{ otherwise}.
		\end{array}
		\right.\nonumber
	\end{align}
	The \kN{} problem is to remove $k$ nodes, i.e., $\sum_{u \in V}x_u~\leq~k$, to break the maximum number of triangles, i.e., to maximize the objective function $\sum_{(u, v, w) \in Tri(V)} y_{uvw}$. Because the triangle $(u, v, w)$ is only broken if at least one node in $\{u, v w\}$ is chosen to be removed, we impose the following constraint,
	\begin{align}
		x_u + x_v + x_w \geq y_{uvw}.\nonumber
	\end{align}
	
	In summary, we have the following equivalent ILP formulation.
	\begin{align}
		\label{eq:kn}
		\max \quad \quad& \sum_{(u, v, w) \in Tri(V)} y_{uvw} \nonumber\\
		\text{s.t.}\quad \quad &\sum_{v \in V}x_v~\leq~k,\nonumber\\
		&x_u + x_v + x_w \geq y_{uvw},\quad \forall (u, v, w) \in Tri(V),\nonumber\\
		&x_u, y_{uvw} \in \{0, 1\}.\nonumber
	\end{align}
	Observe that the above ILP formulation is a special case of the \MC{}\cite{Vazirani01} problem. Given an universe set of elements $\mathcal U$ and a collections of subsets of $\mathcal{U}$, $\mathcal{S} = \{S_1,\dots, S_n\}$ where $S_i \subseteq \mathcal{U}$, the general \MC{} problem asks for $k$ subsets of $\mathcal{S}$, $\hat{\mathcal{S}} = \{\hat S_1,\dots,\hat S_k\}$, to maximize the coverage $\textsf{Cover}(\hat{\mathcal {S}})$ of $\hat{\mathcal{S}}$ where
	\begin{align}
		\textsf{Cover}(\hat{\mathcal {S}}) = \Big|\bigcup_{i = 1}^{k}\hat{S}_k \Big |\nonumber
	\end{align}
	is the number of distinct elements in the union of $\hat{S_i}, i = 1..k$. We call the number of subsets that an element appears in the \textit{frequency} of that element. Thus, in the Eq.~\ref{eq:kn} the universe set is $\mathcal U = Tri(V)$ (i.e. all the triangles) and the collection of subsets is $\mathcal S = \{Tri(v)\ |\ v\in V\}$. This special case of \MC{} also satisfies the condition that  \emph{all the elements have the same frequency three}, as each triangle involves exactly three nodes.

	\begin{definition}[$k$-triangle-breaking-edge]
		Given an undirected graph $G=(V, E)$ and budget size $k$, find a subset $F^*$ of $k$ edges whose removal will break the maximum number of triangles in $G$.
		\begin{align}
			F^* = \argmax &\quad \quad |Tri(F)|\\
			\nonumber\text{s.t.} &\quad \quad |F| \leq k, \\
			\nonumber	&\quad \quad F \subseteq E,
		\end{align}
	\end{definition}
	where $Tri(F)$ is the set of triangles with at least one edge in $F$.
	
	The equivalent ILP of \kE{} is,
	\begin{align}
		\label{eq:ke}
		\max \quad \quad& \sum_{(u, v, w) \in Tri(V)} y_{uvw} \\
		\text{s.t.}\quad \quad &\sum_{(u, v) \in E} x_{uv}~\leq~k,\nonumber\\
		&x_{uv} + x_{vw} + x_{wv} \geq y_{uvw},\quad \forall (u, v, w) \in Tri(V),\nonumber\\
		&x_{uv}, y_{uvw} \in \{0, 1\},\nonumber
	\end{align}
	where \[
	x_{uv} = \left\{ \begin{array}{ll}
	1 & \text{ if edge } (u, v) \text{ is removed,}\\
	0 & \text{ otherwise}.
	\end{array}
	\right.
	\] for all $(u, v) \in E$.
	
	\kE{} is also a special case of \MC{} in which the elements to be covered are the triangles in $G$, and the collection of subsets includes the set of triangles involving each edge $(u, v) \in E$. As each triangle consists of three edges, \emph{the frequency of each element in this instance is also three}. Moreover, any two \emph{subsets have at most one triangle in common}.
	
	We also formulate the converse variants in which we want to break a certain number (or a percentage of the total number) of triangles by removing the least number of nodes/edges from the graph. Their definitions and ILP formulations are defined in the following paragraphs
	
	\begin{definition}[\minN]
			Given an undirected graph $G=(V, E)$ and a positive integer $p \leq |Tri(V)|$, find a minimum-size subset $S$ of \textsf{nodes} whose removal will break at least $p$ triangles in $G$.
		\end{definition}
		\noindent The ILP for \minN is
		\begin{align}
			\label{eq:kn_2}
			\min \quad \quad&  \sum_{v \in V}x_v\\
			\nonumber\text{s.t.}\quad \quad &\sum_{(u, v, w) \in Tri(V)} y_{uvw} \geq p,\\
			\nonumber&x_u + x_v + x_w \geq y_{uvw},\\
			\nonumber&x_u, y_{uvw} \in \{0, 1\}.
		\end{align}
	
	\begin{definition}[\minE]
			Given an undirected graph $G=(V, E)$ and a positive integer $p \leq |Tri(V)|$, find the minimum-size subset $F$ of \textsf{edges} whose removal will break at least $p$ triangles in $G$. 
		\end{definition}
		\noindent The ILP for \minE  is
		\begin{align}
			\label{eq:ke_2}
			\max \quad \quad& \sum_{(u, v) \in E} x_{uv}\\
			\nonumber\text{s.t.}\quad \quad &\sum_{(u, v, w) \in Tri(V)} y_{uvw}\geq p,\\
			\nonumber&x_{uv} + x_{vw} + x_{wv} \geq y_{uvw},\\
			\nonumber&x_{uv}, y_{uvw} \in \{0, 1\}.
		\end{align}
	
	Note that \minN{} and \minE{} are special cases of the \PSC{} problem \cite{Gandhi04}. The \PSC{} problem is a variation of the \textit{set cover} problem. Given an universe set $\mathcal{U}$, a collection of subsets of $\mathcal{U}$, \PSC{} finds a subcollection to cover only a required number $p$ of the elements in $\mathcal{U}$. Thus, \minN{} and \minE{} are equivalent to \PSC{} problems in which each element is in exactly three subsets and the intersection of any three subsets contains at most one element.

%% file: text/npcomplete.tex
\section{Hardness and Approximability}
\label{sshardness}
	We next discuss the complexity and present the best approximation guarantees for our defined problems. The summary of the complexity and approximability results for the studied problems is presented in Table~\ref{tab:comp}.
	
	{\renewcommand{\arraystretch}{1.3}}
	\begin{table*}[t]
		\centering
		\caption{Summary of Complexity and Best Approximation Guarantees}
		\begin{tabular}{llr}
			\addlinespace
			\toprule
			Problem  &  Complexity & Best approximation ratio \\
			\midrule    	
			\kN{} & NP-complete & 19/27 \cite{Ageev04}\\
			\minN{} & NP-complete & 3 \cite{Gandhi04}\\		
			\kE{} & NP-complete &  19/27 \cite{Ageev04}\\		
			\minE{} & NP-complete & 3 \cite{Gandhi04}\\
			\bottomrule		
		\end{tabular}
		\label{tab:comp}
	\end{table*}
	
	\subsection{NP-Completeness}
Recent work of Li et al. \cite{Li14} attempted to prove the NP-completeness of problems similar to \kN{} and \kE{}. Unfortunately, their proofs contained some flaws. Specifically, the proof of Theorem 2.1 \cite{Li14} relies on a weaker constraint of the set system: ``the intersection of any \emph{three} subsets in $\mathcal S$ has at most one element''.
Indeed, for \kE{}, the correct (and stronger) condition should be: the intersection of any \emph{two} subsets in $\mathcal S$ has at most one element.
Moreover, the proof relies on the assumption that if a problem is not NP-hard then there is a polynomial-time algorithm to solve it.
We do not know yet if there exist NP-intermediate problems between NP and P. Consequently, the correctness of the reduction cannot be confirmed.
				
We show that all four aforementioned variants are all NP-complete problems. We present a simple NP-completeness proof of \minN{} (similarly \kN{}) via reduction from the Vertex-Cover problem \cite{Vazirani01}.
The decision versions of \kN{} (similarly \minN{}) can be polynomial-time reducible from the following decision problem, called \emph{Node-Triangle-Free}:
	
	\emph{``Given a undirected graph $G=(V, E)$ and a number $k$, can we delete $k$ nodes from $G$ so that there is no more triangles in $G$ (a.k.a $G$ is triangle-free)?''.}
	
	In turn, we show a more important result that \emph{Node-Triangle-Free} is polynomial-time reducible from the decision version of Vertex Cover problem (definition below). This result will set forth the NP-Completeness of \kN{}. 
	
	\emph{``Given a graph $G=(V, E)$ and an integer $0<k<|V|$, is there a vertex-cover of size $k$?''.}
	
	\textit{\textbf{Reduction:}} Let $\Phi=<G=(V, E), k>$ be an instance of the vertex cover problem. For each edge $(u, v) \in E$, we add to $G$ a new node $t_{uv}$ and connect $t_{uv}$ to both $u$ and $v$. Let $G'$ be the new graph. We shall reduce $\phi$ to an instance $\Lambda = <G', k>$ of \emph{Node-Triangle-Free}.
	Obviously, if we have a vertex-cover $S \subset V$ of size $k$ in $G$ then we can delete the same set of nodes $S$ in $G'$ to obtain a triangle-free graph. In the reverse direction, we can assume without lost of generality that $t_{uv}$  will never be removed. The reason is that we can always remove $u$ or $v$ and break an equal or greater number of triangle(s). Thus a subset of size $k$ that its removal makes $G'$ triangle-free must induce a vertex-cover of size $k$ in $G$. This completes the reduction. 
	
	\begin{theorem}
		The problems \kN {} and \minN{} are NP-complete.
	\end{theorem}
	
	Using a similar reduction, both \kE{} and \minE{} can be polynomial-time reducible to the following problem:
	
	\emph{``Can we delete $k$ edges from a graph $G=(V, E)$ so that there is no more triangles in $G$ (i.e., to make the graph triangle-free)?''.}
	
	The above problem is known to be NP-complete according to \cite{Yannakakis81}. Hence, we obtain the following result. 
	
	\begin{theorem}
		The problems \kE{} and  \minE{} are NP-complete.
	\end{theorem}
		
	\subsection{Approximability} 
	Since \minN{} and \minE{} problems are special cases of the \PSC{} problem with bounded frequencies $f=3$ \cite{Gandhi04}, the primal-dual algorithm in \cite{Gandhi04} provides a 3-approximation algorithm for both problems. 
Instead of operating on sets, the primal-dual algorithm works on the elements in the universe set $\mathcal{U}$.
It assigns a dual covering cost for each element that signifies the selection of a set to cover that element.
The basic operation of the algorithm is increasing all the dual covering costs of those that have not been covered simultaneously until the total cost of uncovered elements in a set equals 1 (the cost of choosing that set).
The corresponding set is then selected to the solution and the algorithm continues until satisfying the covering requirement.
To achieve the $f$-approximation factor, the algorithm assumes that we know a set in the optimal solution (simply by trying all the possible sets) and applies the primal-dual selection on the rest. Therefore, we obtain the following result.
	
	\begin{theorem}
		There exist 3-approximation algorithms for \minN{} and \minE.
	\end{theorem}
	
	The \kN{} and \kE{} problems are special cases of \MC{} and the Pipage-rounding method in \cite{Ageev04} results in an approximation algorithm with ratio $1-(1-1/3)^3 = 19/27$. 
	
	The Pipage-rounding is a general method providing worst-case approximation guarantees for a large class of discrete optimization problems, including \MC{}, with assignment-type constraints. It first reformulates the problem into a non-linear program which has an integral optimum and is at least $1-(1-1/f)^f$ greater than the starting problem at any feasible solution. It then finds an integral solution of the non-linear program in two phases: 1) solving the non-integral relaxation of the problem and 2) transform the non-integral solution to an integral one by pipage rounding. The relaxation is polynomially solvable and the second phase takes the solution and rounds it in the manner that the objective value of rounded solution can only increase and get closer to integral numbers. As shown in \cite{Ageev04}, each rounding circle in Pipage-rounding brings one element in the current solution to integral value. The approximation factor follows directly from the properties of the non-linear program and the rounding procedure. Therefore, we obtain the following result.
	
	\begin{theorem}
		There exist 19/27-approximation algorithms for \kN{} and \kE.
	\end{theorem}

\textit{\textbf{Remarks}}. Both the primal-dual method in \cite{Gandhi04} and the pipage-rounding algorithm in \cite{Ageev04} have high time complexity and are not scalable for large networks.
As a result, efficient algorithms that can be applied on large-scale data are of desire. In next sessions, we propose efficient discounting algorithms for the studied problems on very large-scale networks with just a slightly looser approximation ratio.

%% file: text/algorithms.tex
\section{Algorithms for \kN{}}
\label{sec:algs_node}
	In this section, we first present a naive Greedy Algorithm (Alg.~\ref{algo:greedy}) to solve the \kN{} problem. We show that the greedy strategy returns an $(1-1/e)$- approximate solution but has prohibitively high time complexity. Thus, in the subsequent subsection, we propose \kN{} Discounting Algorithm (\IN{} - Alg.~\ref{algo:fast_greedy}) which achieves the same solution quality but is at least $k$ time faster. The core efficiency of \IN{} is that it employs a smart updating technique to keep track of the number of effective triangles associated with each of the remaining nodes.
	
	\begin{algorithm}
		\caption{Greedy Algorithm for \kN{} (Simple\_Greedy)}
		\label{algo:greedy}
		\begin{algorithmic}[1]
			\State $S \gets \emptyset$;
			\State \textbf{for} i = 1 \textbf{to} k
			\State \quad $S \leftarrow S + \arg\max_{v \in V\setminus S} \Delta_S(v)$;    			
			\State \Return{$S$}
		\end{algorithmic}
	\end{algorithm}
	\subsection{Naive Greedy Algorithm}
	
	The first algorithm (Alg.~\ref{algo:greedy}) selects at each step the node $u$ that breaks the most number of triangles, i.e., $u = \arg\max_{v \in V\setminus S} \Delta_S(v)$, and then adds $u$ to the solution $S$. This algorithm continues until $k$ nodes have been selected into the returned solution $S$.
	
	Since \kN{} is a special case of \MC{}, the native greedy algorithm provides a performance guarantee of $(1-1/e)$ for \kN{}. Another way of proving this is to show that the main objective function (the number of broken triangles) is monotone and submodular, which in turn admits a nearly optimal greedy approximation algorithm \cite{Li14}.
		
	The complexity of Alg.~\ref{algo:greedy} is $O(k m n)$ assuming $k$ nodes are selected in the solution. In a recent work, the time complexity for Alg. \ref{algo:greedy} is brought down to $O(k m^{3/2})$ in \cite{Li14} using the fast triangle computation method in \cite{Schank05}. For large value of $k=\theta(n)$, the time-complexity of the algorithm in \cite{Li14} could be as high as $O(n m^{3/2})$ which is very expensive and not scalable for practical large size data. To this end, we present in next section our scalable Discounting Algorithms for \kN{} with time complexity $O(m^{3/2} + km)$ which is up to $m^{1/2}$ times faster than the algorithm in \cite{Li14}.
	
	\begin{algorithm}[t]
		\caption{Discounting Algorithm for \kN{} (\IN{})}
		\label{algo:fast_greedy}
		\begin{algorithmic}[1]
			\vspace{0.05in}
			\Statex \textbf{Phase 1:}
			\vspace{0.05in}
			\State Number nodes from $1$ to $n$ such that $u < v$ implies $d(u) \leq d(v)$.
			\State $S \gets \emptyset$;
			\State {\bf for  each} $u \in V$ {\bf do} $T(u) \gets 0$;
			\State {\bf for} $u \gets n \textrm{ to } 1$ {\bf do} 
			\State \qquad {\bf for  each} $v \in N(u)$ with $v<u$ {\bf do} 
			\State \qquad \qquad {\bf for  each}  $w \in A(u) \cap A(v)$ {\bf do} 
			\State \qquad \qquad \qquad Increase $T(u), T(v)$ and $T(w)$ by  one;
			\State \qquad \qquad Add $u$ to $A(v)$;
			\vspace{0.1in}
			\Statex \textbf{Phase 2:}
			\vspace{0.05in}
			\State $Q \leftarrow $Max-Priority-Queue$(T)$
			\State {\bf for} i = 1 \textbf{to} k
			\State \qquad $u_{max} = Q.pop()$;
			\State \qquad Remove $u_{max}$ from $G$ and add $u_{max}$ to $S$;      	
			\State \qquad {\bf for  each} $ v \in N(u_{max})$ {\bf do}
			\State \qquad \qquad {\bf for  each} $ w \in N(v)$ {\bf do}
			\State \qquad \qquad \qquad {\bf if} $ v, w \in N(u_{max}) \setminus S$ {\bf then} 
			\State \qquad \qquad \qquad \qquad Decrease $T(v)$ and $T(w)$ by one;
			\State \qquad \qquad \qquad \qquad $Q.update(v,T)$;
			\State \qquad \qquad \qquad \qquad $Q.update(w,T)$;  	
			\State \Return{$S$}			
		\end{algorithmic}	
	\end{algorithm}
	
	\subsection{Discounting Algorithm for \kN{}}

	Our Discounting Algorithm for \kN{} (\IN{} - Alg. \ref{algo:fast_greedy}) speeds up significantly the simple greedy algorithm. For small values of $k$, this algorithm requires as much time as the best algorithm for counting the number of triangles.
	
	In principle, \IN{} employs an adaptive strategy in computing the marginal gains (the number of broken triangles) when nodes are removed one after another. At each round, the node $v$ that breaks the most number of triangles is selected into the solution. Node $v$ is then excluded from the structure and the procedure repeats itself on the remaining nodes and  \emph{recomputes}  efficiently the new marginal gain for each node $u$. 
	
	We structure \IN{} into two phases. The first phases (lines 1--8) extends the algorithm in \cite{Schank05} to compute the number of triangles that are incident with each node in the graph. This algorithm was proved to be time-optimal in $\theta(m^{3/2})$ for triangle-listing, and has been shown to be very efficient in practice. The second phase starts at line 9 where it creates a \textit{Max-priority-queue} to ranks nodes according to values in $T$. \IN{} then (lines 9--18) repeats the vertex selection for $k$ rounds. In each round, we select the node $u_{max}$ with the highest value of $\Delta_S(u)=T(u)$ (from top of the priority queue) into the solution. The algorithm then removes $u_{max}$ from the graph, and performs the necessary updates on $T(u)$ for all $u \in V \setminus S$. The algorithm subsequently updates the positions of the nodes $v$ and $w$ in the queue according to the new values of those nodes in $T$.
	The key efficiency of \IN{} algorithm lies in its update procedure for $\Delta_S(u)=T(u)$. Specifically, the total update for all $O(n)$ values of $\Delta_S(u)$ after removing $u_{max}$ can be done in \textit{linear time} as indicates in lines 15 -- 18. The linear time update is made possible due to the information on the number of triangles involving each node. This significantly reduces the complexity for computing the marginal gain $\Delta_S(u)$ and speeds up the node selection process.
	
	
	
	\textit{Complexity}: The first phase takes $O(m^{3/2})$ as in \cite{Schank05}. The second phase takes a linear time in each round and has a total time complexity $O(k(m+n))$ as creating and maintaining the Max-priority queue requires $O(n\log n)$. In each sequential round, the algorithm checks all the neighbors $v$ of $u_{max}$ and for each neighbor, it examines all the neighbors of $v$. Thus, the total complexity of checking at a round is $\sum_{v\in N(u_{max})} d_v \leq 2m$ where $d_v$ is the degree of $v$. Each update (Lines 17-18) takes constant time since $T(v)$ and $T(w)$ decrease by 1 and the queue $Q$ needs to move $v,w$ at most one level in the queue. Thus, the overall complexity is $O(m^{3/2} + km )$. For $k < m^{1/2}$, the algorithm has an effective time-complexity $O(m^{3/2})$, which is the same as the counting triangles procedure.
	
	\textit{Approximation guarantees:} It is obvious that \IN{} respects the original greedy method as it selects the node with the highest marginal gain at each step. Hence, \IN{} retains the approximation guarantees of the greedy method for \MC{}. The following theorem summarizes our suggested approach.
	\begin{theorem}
		\IN{} algorithm is an $(1-1/e)$-approximation algorithm for \kN{} with complexity $O(m^{3/2} + km)$.
	\end{theorem}
	
	Note that the naive Greedy (Alg.~\ref{algo:greedy}) and Discounting Algorithms (Alg.~\ref{algo:fast_greedy}) can be easily adapted for \minN{} by stopping selecting nodes until $p$ broken triples triangles are satisfied. This is due to the fact that \minN{} is a special case of the \PSC{} problem and the greedy strategy guarantees an $H(p)-1/2$ approximation solution, where $H(p)$ denotes the harmonic function $H(p)=1+1/2+\ldots+1/p$. Thus, Algs.~\ref{algo:greedy} and~\ref{algo:fast_greedy} are ($H(p)-1/2$)-approximation algorithms for \minN{}.
	
	\subsection{Analysis in Networks with Power-law Degree Distribution}
	
As discussed above, \IN{}'s time complexity is $O(m^{3/2} + km )$ for a general network; however, many complex systems of interest such as the Internet, social, and biological networks commonly exhibit the power-law degree distributions \cite{Barabasi00,Barabasi02}. 
Conceptually, power-law degree distributed networks have the fraction of nodes with degree $k$ ($k$ connections to other nodes) is $\lfloor{\frac{e^{\alpha}}{k^{\gamma}}}\rfloor$, where $e^{\alpha}$ is the normalization factor as in the $P(\alpha,\gamma)$	model \cite{Aiello01}.
Practical networks usually have $2 < \alpha < 3$.
In this work, we deduce the maximum degree in a $P(\alpha,\gamma)$ network to $e^{\frac{\alpha}{\gamma}}$ because for $k>e^{\frac{\alpha}{\gamma}}$, the number of edges will be less than 1.
We show that in power-law degree distributed networks, the overall time complexity is $O(m^{3/2})$ which implies that \IN{} is as fast as the state-of-the-art algorithms for counting/listing triangles with no additional costs (Theorem \ref{thr:dakn}. This also realizes the scalability of \IN{} in large networks. 
	\DeclarePairedDelimiter\ceil{\lceil}{\rceil}
	\DeclarePairedDelimiter\floor{\lfloor}{\rfloor}
	
	\begin{theorem}
		The complexity of \IN{} algorithm is $O(m^{\frac{3}{2}})$ on power-law degree distributed networks. This implies \IN{} is as fast as the best available triangle counting/listing algorithms.
		\label{thr:dakn}
	\end{theorem}
	
	\textbf{\textit{Proof: }}		
In a power-law degree distributed network, the numbers of vertices and edges are computed as follows,
		\begin{align}
			\label{eq:nodes}
			n = \sum_{k=1}^{e^{\frac{\alpha}{\gamma}}} \frac{e^{\alpha}}{k^{\gamma}} \approx \left\{ \begin{array}{lll}
			\zeta(\gamma)e^{\alpha} & \text{ if }\gamma > 1\\
			\alpha e^{\alpha} & \text{ if } \gamma = 1 ,\\
			\frac{e^{\frac{\alpha}{\gamma}}}{1-\gamma} & \text{ if } \gamma < 1
			\end{array}
			\right.
		\end{align}
		\begin{align}
			\label{eq:edges}
			m = \frac{1}{2}\sum_{k=1}^{e^{\frac{\alpha}{\gamma}}} k \frac{e^{\alpha}}{k^{\gamma}} \approx \left\{ \begin{array}{lll}
			\frac{1}{2}\zeta(\gamma-1)e^{\alpha} & \text{ if }\gamma > 2\\
			\frac{1}{4}\alpha e^{\alpha} & \text{ if } \gamma = 2\\
			\frac{1}{2}\frac{e^{\frac{2\alpha}{\gamma}}}{2-\gamma} & \text{ if } \gamma < 2
			\end{array}
			\right.
		\end{align}
		where $\zeta(\gamma) = \sum_{i = 1}^{\infty} \frac{1}{i^\gamma}$ is the Riemann Zeta function \cite{Aiello01,Dinh132} which converges absolutely for $\gamma > 1$ and diverges for all $\gamma \leq 1$.
		For the sake of simplicity, we will	simply use real number instead of rounding down to integers. The error terms can be easily bounded and are negligible in our proof.
		
		Since Phase 1 of Alg.~\ref{algo:fast_greedy} is $O(m^{\frac{3}{2}})$ for counting triangles, we will analyze phase 2 in Alg.~\ref{algo:fast_greedy} and show its complexity $O(m^{\frac{3}{2}})$. To this end, we first find the workload $C_i$ at each round $i$ in phase 2, sum them all up and utilize the power-law property to obtain the final result. In particular,
		\begin{align}
			C_i = \sum_{v \in N(u_{max})} d_v \nonumber
		\end{align}
		
		The worst case of the second phase happens when $k = n$ which means that the algorithm has to select all nodes in decreasing order of triangle-breaking gains into the solution set $S$. That leads to the overall complexity of,
		\begin{align}
			\label{eq:complexity}
			C = \sum_{i = 1}^{n} C_i = \sum_{u \in V}\sum_{v \in N(u)} d_v = \sum_{u \in V} d^2_u
		\end{align}
		
		We apply the power-law property on the number of nodes with degree $k$ being $\frac{e^{\alpha}}{k^{\gamma}}$ and the maximum degree is $e^{\frac{\alpha}{\gamma}}$ on the above equation which yields
		\begin{align}
			\label{eq:com}
			C = \sum_{u \in V} d^2_u = \sum_{k = 1}^{e^{\frac{\alpha}{\gamma}}} k^2 \frac{e^{\alpha}}{k^{\gamma}} = e^{\alpha} \sum_{k = 1}^{e^{\frac{\alpha}{\gamma}}} k^{2-\gamma}
		\end{align}
		
		We consider two cases:
		
		\textbf{Case 1: $\gamma \geq 2$.} This implies $k^{2-\gamma} \geq 1$. Eq.~\ref{eq:com} becomes,
		\begin{align}
			\label{eq:com1}
			C = e^{\alpha} \sum_{k = 1}^{e^{\frac{\alpha}{\gamma}}}k^{2-\gamma} &\leq e^{\alpha} \sum_{k = 1}^{e^{\frac{\alpha}{\gamma}}}1 = e^{\alpha} e^{\frac{\alpha}{\gamma}} = e^{\alpha + \frac{\alpha}{\gamma}} \nonumber\\
			&\leq e^{\alpha + \frac{\alpha}{2}} = \Big( e^{\alpha}\Big)^{\frac{3}{2}}.
		\end{align}
		
		Combining Eq.~\ref{eq:com1} with the number of edges in power-law degree networks in Eq.~\ref{eq:edges}, we obtain,
		\begin{align}
			\label{eq:com2}
			C \leq \Big( e^{\alpha}\Big)^{\frac{3}{2}} = \textbf{c1} \cdot m^{\frac{3}{2}}.
		\end{align}
		where $\textbf{c1}$ is a constant that satisfies,
		\begin{align}
			\textbf{c1} \approx \left\{ \begin{array}{ll}
			(\frac{1}{\frac{1}{2}\zeta(\gamma-1)})^{3/2} & \text{ if }\gamma > 2.\\
			(4/\alpha)^{3/2} & \text{ if } \gamma = 2.
			\end{array} \right.\nonumber
		\end{align}
		Note that $\gamma > 2$ infers $\zeta(\gamma-1)$ converges and $\textbf{c1}$ is a finite constant.
		
		Thus, in this case, phase 2 has time complexity of $O(m^{\frac{3}{2}})$.
		
		\textbf{Case 2: $\gamma < 2$}. In this case, Eq.~\ref{eq:com} is equivalent to,
		\begin{align}
			\label{eq:com3}
			C &= e^{\alpha} \sum_{k = 1}^{e^{\frac{\alpha}{\gamma}}} k^{2-\gamma} = e^{\alpha}(e^{\frac{\alpha}{\gamma}})^{2-\gamma} \sum_{k = 1}^{e^{\frac{\alpha}{\gamma}}} \frac{k^{2-\gamma}}{(e^{\frac{\alpha}{\gamma}})^{2-\gamma}}\nonumber\\
			&\leq e^{\alpha}e^{\frac{2\alpha}{\gamma}-\alpha} \int_{t = 0}^{1} t^{2-\gamma} \text{d}t = e^{\frac{2\alpha}{\gamma}}\frac{1}{3-\gamma} = \textbf{c2} \cdot m.
		\end{align}
		where 
			$$\textbf{c2} \approx \frac{2 (2-\gamma)}{3-\gamma},$$
		is a finite constant since $\gamma < 2$. This yields the time complexity $O(m)$ for Phase 2. Finally, we conclude that the overall time complexity of $O(m^{\frac{3}{2}})$ in both cases.
	
	\section{Algorithm for \kE{}}
	\label{sec:algs_edge}

	\begin{algorithm}[h]
		\caption{Discounting Algorithm for \kE{} (\IE{})}
		\label{algo:fast_greedy_e}
		\begin{algorithmic}[1]
			\vspace{0.05in}
			\Statex \textbf{Phase 1:}
			\vspace{0.05in}
			\State Renumber nodes so that $u < v$ implies $d(u) \leq d(v)$.
			\State $F \gets \emptyset$;   
			\State {\bf for  each} $(u, v) \in E$ {\bf do} $tr(u, v) \gets 0$;
			\State {\bf for} $u \gets n \textrm{ to } 1$ {\bf do} 
			\State \qquad {\bf for  each} $v \in N(u)$ with $v<u$ {\bf do} 
			\State \qquad \qquad {\bf for  each}  $w \in A(u) \cap A(v)$ {\bf do} 
			\State \qquad \qquad \qquad Increase $tr(u, v), tr(v, w)$ and $tr(u, w)$ by one;
			\State \qquad \qquad Add $u$ to $A(v)$;
			\vspace{0.1in}
			\Statex \textbf{Phase 2:}
			\vspace{0.05in}
			\State $Q\leftarrow$ Max-Priority-Queue$(T)$
			\State {\bf For} i = 1 \textbf{to} k
			\State \qquad $e_{max} \gets Q.pop()$;
			\State \qquad Remove $e_{max}$ from $G$ and add $e_{max}$ to $F$;      	
			\State \qquad Let $(u', v') = e_{max}$;
			\State \qquad {\bf for  each} $w \in N(u') \cap N(v')$ {\bf do}       
			\State \qquad \qquad Decrease $tr(w, u')$ and $tr(w, v')$ by one;      	
			\State \qquad \qquad $Q.update((w,u'),T)$;
			\State \qquad \qquad $Q.update((w,v'),T)$;
			\State \Return{$F$}			
		\end{algorithmic}	
	\end{algorithm}
	
Similarly to \kN{} and \minN, the edge variants expose similar attributes and thus the greedy algorithm can be directly applied with near-optimal guarantee. We present \IE{} for finding triangle-breaking edges in Alg.~\ref{algo:fast_greedy_e}. On general networks, \IE{} is faster than its node-version, \IN{}, because it possesses a complexity $O(m^{3/2} + kn)$.

Unlike \IN{}, \IE{} maintains for each edge the number of triangles incident on that edge and updates the measure efficiently when removing nodes from $G$. After removing an edge $(u', v')$ we only needs to consider only  $|N(u') \cap N(v')|$ updates to discount the triangles incident on $(u', v')$ from the corresponding edges. Thus the overall complexity in each iteration relies on finding the edge that breaks the maximum number of triangles.
Similar to the node version, we also have the same approximation guarantees for the edge-removal problems which is summarized below.
	\begin{theorem}
		\IE{} is an $(1-1/e)$-approximation algorithm for \kE{} with complexity $O(m^{3/2} + kn)$.
	\end{theorem}
	
	On power-law degree distributed networks, by similar arguments to \IN{}, we can show that the overall complexity of \IE{} is $O(m^{\frac{3}{2}})$ which is also equal to that of counting/listing triangles in the networks.
	\begin{theorem}
		On power-law degree distributed networks, the complexity of \IE{} algorithm is $O(m^{\frac{3}{2}})$.
	\end{theorem}
	
	An easily adapted algorithm of Alg.~\ref{algo:fast_greedy_e} can be devised for solving \minE{} and returns a ($H(p)-1/2$)-approximate edge set since \minE{} is also a special case of \PSC{} problem.
	
	\subsection*{Input-dependent approximation guarantees}
	
	The $(1-1/e)$-approximation factor, termed \textit{fixed worst-case bound}, achieved by our algorithms provides a general lower-bound on the solution quality of the selected set $S$. This factor is known in advance even prior to the execution of the methods. Nevertheless, we can often times derive a better approximation bound of the solution quality, namely the \textit{input-dependent bound}, depending on the problem instance and even the particular run of the algorithms.
		Inspired by the work in \cite{Leskovec07} on the Influence Maximization problem, we can apply a similar bounding technique (named online-bound) to obtain a real input-dependent bound on the solution quality in both the naive greedy and our \IN{} and \IE{} algorithms. The input-dependent bound for \IN{} is stated as follows,
	\begin{theorem}[\IN{} input-dependent bound]
		For a set of selected nodes $S \subset V$ and each node $u \in V$, let $\Delta_S(u) = T(S\cup{u}) - T(S)$ be the marginal gain of $u$ when $u$ is included in $S$. Let $u_1,u_2,\dots,u_{n-k}$ be the sequence of the remaining nodes (not in $S$) sorted in decreasing order of $\Delta_S(u)$, then
		\begin{align}
			OPT^n_k \le T(S) + \sum_{i = 1}^{k} \Delta_S(u_i)
		\end{align}
		where $OPT^n_k = \max_{S' \subset V, |S'| = k} T(S')$ is the triangles broken by the optimal solution with $k$ nodes.
	\end{theorem}
	
	By selecting the top $k$ nodes with largest marginal triangle-breaking gains into the returned solution $S$ of \IN{}, we obtain an upper-bound on the optimal solution. Then by dividing the number of triangles broken by $S$ with that upper-bound, we have an input-dependent guarantee on $S$,
	\begin{align}
		\label{eq:ob_n}
		\mathcal{OB}_n(S) = \frac{T(S)}{T(S) + \sum_{i = 1}^{k}\Delta_S(u_i)} \ge \frac{T(S)}{OPT^n_k}
	\end{align}
	
	Similarly, the input-dependent for solution $F$ of the \IE{} is computed by the following equation,
	\begin{align}
		\label{eq:ob_e}
		\mathcal{OB}_e(F) = \frac{T(F)}{T(F) + \sum_{i = 1}^{k}\Delta_F(e_i)} \ge \frac{T(F)}{OPT^e_k}
	\end{align}
	where $e_1,\dots,e_k$ are the top $k$ edges with the highest marginal gain of broken triangles with respect to $F$ and $OPT^e_k$ is the triangles broken by the optimal edge set with $k$ edges.

%% file: text/experiments.tex
\section{Experimental Evaluation}
\label{ssexperiments}

\setlength\tabcolsep{3pt}
\begin{table*}[t]
	\caption{Real-world networks for experimentation}
	\label{tab:data}
	\centering
	\begin{tabular}{ l l r  r  c }\toprule
		\textbf{Dataset} & \textbf{	Type} & \bf \#Nodes& \bf \#Edges & \bf Avg. degree\\
		\midrule
		Gnutella4 & Peer-to-peer network\textsuperscript{(*)} & 10.9K & 40K & 3.7\\
		Flickr & Photo sharing network\textsuperscript{(\textdagger)} & 80.5K & 11.8M & 138.8\\
		Google & Web graph\textsuperscript{(*)} & 876K & 5.1 M & 5.83 \\
		Skitter & Internet Topology\textsuperscript{(*)} & 1.7M & 11.1M & 6.53\\
		Wiki-Talk & Wikipedia Communication\textsuperscript{(*)} & 2.4M & 5M & 2.1\\
		Orkut & Online Social Network\textsuperscript{(*)} & 3M & 117M & 78\\
		\bottomrule
	\end{tabular}
\end{table*}

In this section, we evaluate the quality and performance of our proposed methods, i.e., \IN{} and \IE{}. Empirical results show two important features of our approaches: \textit{performance} and \textit{scalability} that are desired for any
practical techniques. We compare and contrast ours with the state-of-the-art method, GreedyAll \cite{Li14} \footnote{\cite{Li14} also proposed another algorithm, namely \textsf{Approx} which used FM-sketch to approximate the triangle-breaking gain; however, this approximation algorithm imposes the same time complexity with GreedyAll.}, and approaches based on centrality measures, i.e., Max-degree, Pagerank and randomization. 
On \kN{} and \kE{}, results indicate that our methods vastly outperform GreedyAll up to orders of magnitudes in terms of running time while achieving the same level of solution quality. The baseline methods based on centrality and randomization are slightly faster but the qualities are much worst. We also spend a good portion to study the networks under node and edge removal attacks using the \minN{} and \minE{}.

\subsection{Experimental settings}

\subsubsection*{Datasets}
To make our experiments extensive, we select a set of six real-world traces from various domains with sizes ranging from thousand to million scales.
The summary of those networks are provided in Table.~\ref{tab:data}.
\blfootnote{\textsuperscript{(*)} http://snap.stanford.edu/data/index.html;\\\indent\textsuperscript{(\textdagger)} http://socialcomputing.asu.edu/pages/datasets}

Specifically, our dataset includes both physical (connected by physical links) and virtual (e.g., friendship, communication) networks. In the first category: Gnutella4 is a snapshot of the Gnutella peer-to-peer file sharing network on August 4th 2002 in which nodes represent hosts in the Gnutella network topology and edges represent connections between the hosts; Skitter is the Internet topology graph captured by tracerouting in 2005. In the second category: Flickr is a contact network crawled from the photo sharing Flickr website where nodes are users and edges are friendship connections between users; Google is the dataset of webpages and hyperlinks between the webs released by Google company in 2002; Wiki-Talk contains the set of users in the Wikipedia website and edit relationship (who edits take pages of whom) and Orkut is an online social networks with users as nodes and friendships as connections.

\subsubsection*{Performance and Scalability measures}

(Performance) For a fair comparison between different methods, we count the number of triangles broken by the set of nodes/edges returned by the algorithms as the quality measure.

\noindent(Scalability) In terms of scalability, we record the running time consumed by each algorithm. For the \minN{} and \minE{} problem, we only measure the running time of \IN{} and \IE{}. The input-dependent bound of our algorithms is also illustrated in the last experiments.

\subsubsection*{Implementation and Testing Environment}

We implemented our algorithms \IN{} and \IE{} in C++ programming language with GCC 4.8 C++11 compiler. We also implemented the GreedyAll \cite{Li14} algorithm following closely the provided description and pseudo-code. All the experiments are run on a Linux environment with 2.2Ghz Xeon 8 core processor and 100GB of RAM. In each execution, only a single core is assigned for each method.

\subsection{Performance Evaluation}
The performance, i.e., solution quality, measured by the number of triangles broken by the node or edge sets returned by the algorithms is illustrated in Figs.~\ref{fig:node_tri} and~\ref{fig:edge_tri} for node and edge variants, respectively.
As depicted from these figures, \IN{}, \IE{} and GreedyAll consistently have the best performance on all the social traces compared to the others.
Pagerank and Max-degree achieve very good solution quality on certain datasets, e.g., Google and Wiki-Talk, but fall far behind \IN{}, \IE{} and GreedyAll on the other tests.
The quality of Random strategy, as expected, falls below and is inconsistent compared to the others.
In summary, empirical results from multiple real-world data confirm the performance provided by our suggested algorithms. 

Figs.~\ref{fig:node_tri} and~\ref{fig:edge_tri} also display the typical trend of monotone and submodular functions as they exhibit the diminishing return property. For the first few selections, the marginal gain (in terms of the number of broken triangles) is significant yet the later rounds provide smaller marginal gain, and the gain tends to saturate quickly.

\subsection{Scalability Evaluation}
Figs.~\ref{fig:node_time} and~\ref{fig:edge_time} report the time consumption (in seconds) of testing algorithms in experiments.
These figures display three groups of methods with different magnitudes: (1) GreedyAll with most time consumption (up to 100x times higher than the second group) (2) \IN{}, \IE{}, Pagerank and Max-degree algorithms, and (3) Random method which returns almost instantly $k$ random nodes/edges.
Our suggested algorithms \IN{} and \IE{} require comparable amount of time as Pagerank and Max-degree which are two canonical centrality measures and very fast to compute. Better yet, \IN{} and \IE{} produce much better solution quality than Pagerank and Max-degree while are very comparable in terms of scalability.

These extensive experiments illustrate that our proposed \IN{} and \IE{} algorithms is highly competitive to the current best GreedyAll method performance meanwhile is much better in terms of scalability. As shown in the previous experiments, only GreedyAll has similarly highest level of solution quality as \IN{} and \IE{}; however, our running time results show that GreedyAll is up to 20 slower than \IN{} on the node removal problem and 100 times slower than \IE{} on the edge removal variants.

\setlength\tabcolsep{5pt}
\begin{table*}[t]
	\caption{Input-dependent bounds provided by \IN{} (closer to 1 is better)}
	\label{tab:ob}
	\centering
	\begin{tabular}{ l ccccc}\toprule
		\textbf{Data} & $k = 200$ & $k = 400$ & $k = 600$ & $k = 800$ & $k = 1000$\\
		\midrule
		Flickr  & 0.65 & 0.74 & 0.81 & 0.85 & 0.88\\
		Gnutella & 0.77 & 0.90 & 1 & 1 & 1 \\
		Google & 0.78 & 0.78 & 0.78 & 0.79 & 0.79\\
		Skitter & 0.77 & 0.80 & 0.82 & 0.84 & 0.85\\
		Wiki-Talk & 0.84 & 0.95 & 0.97 & 0.99 & 0.99\\
		Orkut & 0.75 & 0.79 & 0.81 & 0.81 & 0.82\\
		\midrule
	\end{tabular}
\end{table*}
\subsection{Input-dependent bound testing}

Finally, we perform  experiments on the input-dependent bounding technique embedded in \IN{} and \IE{} algorithms. Theoretically, the solutions returned by \IN{} and \IE{} are guaranteed to be at least $(1-1/e) \approx 0.63$ on any problem instance. In practice, we can have better guarantee depending on the problem instance and the execution itself. Our input-dependent bounding strategy is one way of finding such instance- and execution-dependent guarantees.

Table~\ref{tab:ob} presents the input-dependent bounds provided by our proposed \IN{} algorithm for node removal problem. This table shows the input-dependent bounds are substantially better than the theoretical guarantee $1-1/e \approx 0.63$. For example, with $k=400$ on Wiki-Talk, \IN{} guarantees solution at 95\% optimal. For the case of Gnutella network, with $k \ge 600$, \IN{} guarantees to find the optimal solution, implying that all the triangles have been disrupted. One can also observe that the bound gets tighter when $k$ increases. This is explainable due to the nature of our bounding technique: larger $k$ means more triangles are broken and the gain of the next $k$ nodes becomes smaller and approximation ratio approaches 1.

%% file: text/conclusion.tex
\section{Conclusion}
\label{ssconclusion}
In this paper, we study the problems of finding critical nodes and links whose failures will severely damage most triangles in the network, changing the network's organization and (possibly) leading to the unpredictable dissolving of the network.
We formulate this vulnerability analysis as optimization problems, and provide proofs of their NP-Completeness. We propose two algorithms \IN{} and \IE{} with notable performance and scalability. Both \IN{} and \IE{} obtain best approximation guarantees: 19/27-approximation for \kN{} and \kE{} as well as 3-approximation for \minN{} and \minE, and are scalable for network with millions nodes and edges. Those features lend our approaches nicely into the analysis of various large-scale real-world problems. In the future, we aim to bridge the gaps between theory and practice to design the scalable approximation with best possible approximation ratios.

%% file: text/expfigures.tex
\begin{figure*}[ht]
	\centering
	\subfloat[Flickr]{\includegraphics[width=0.33\textwidth]{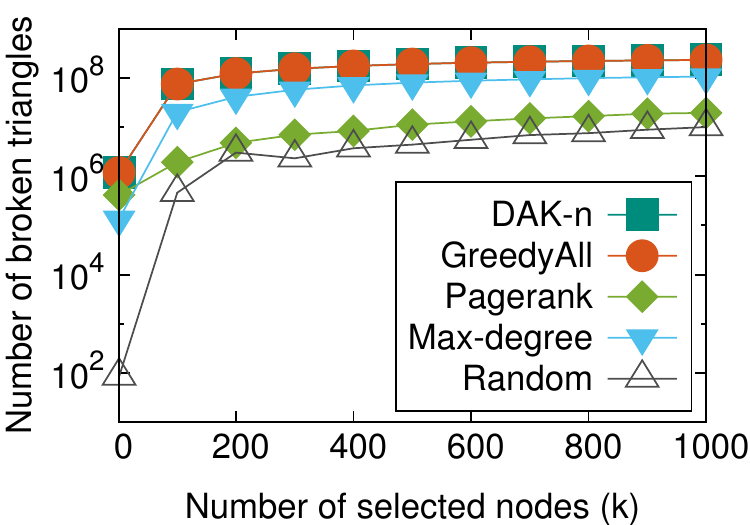}}
	\subfloat[Gnutella]{\includegraphics[width=0.33\textwidth]{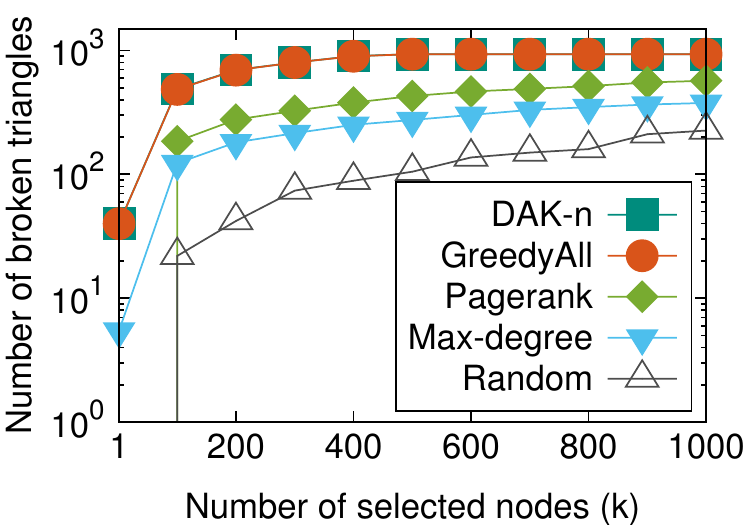}}\subfloat[Google]{\includegraphics[width=0.33\textwidth]{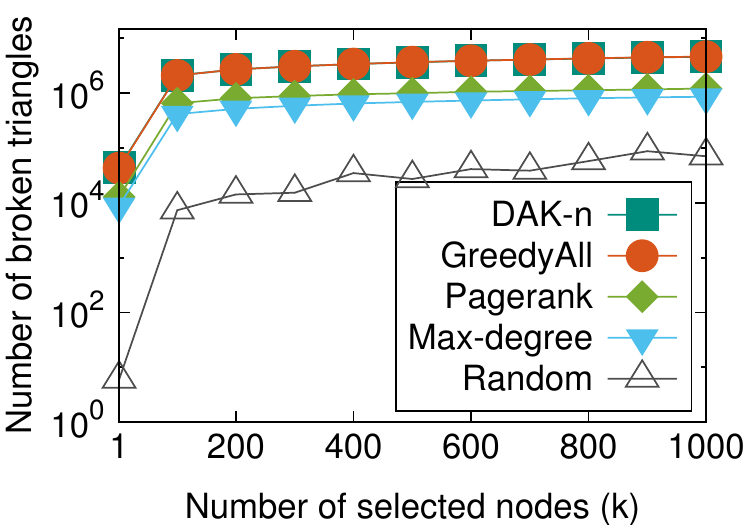}}\vspace{-0.1in}\
	\subfloat[Skitter]{\includegraphics[width=0.33\textwidth]{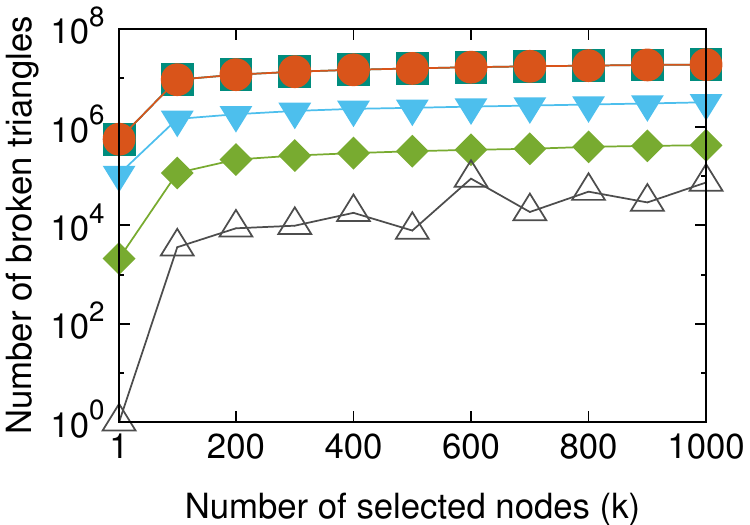}}
	\subfloat[Wiki-Talk]{\includegraphics[width=0.33\textwidth]{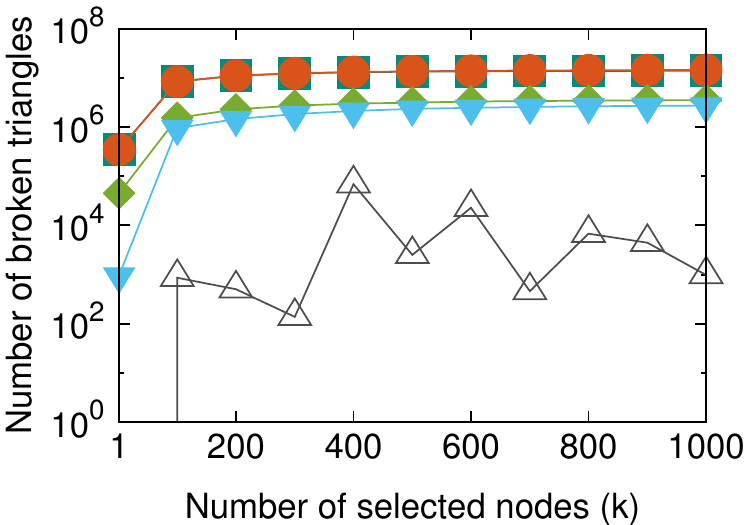}}
	\subfloat[Orkut]{\includegraphics[width=0.33\textwidth]{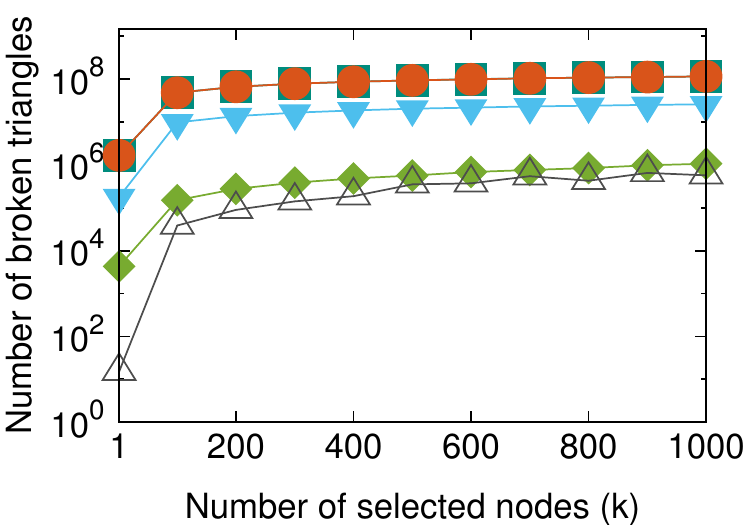}}
	\caption{Number of broken triangles by node removal (higher value is better)}
	\label{fig:node_tri}
\end{figure*}

\begin{figure*}[ht]
	\centering
	\subfloat[Flickr]{\includegraphics[width=0.33\textwidth]{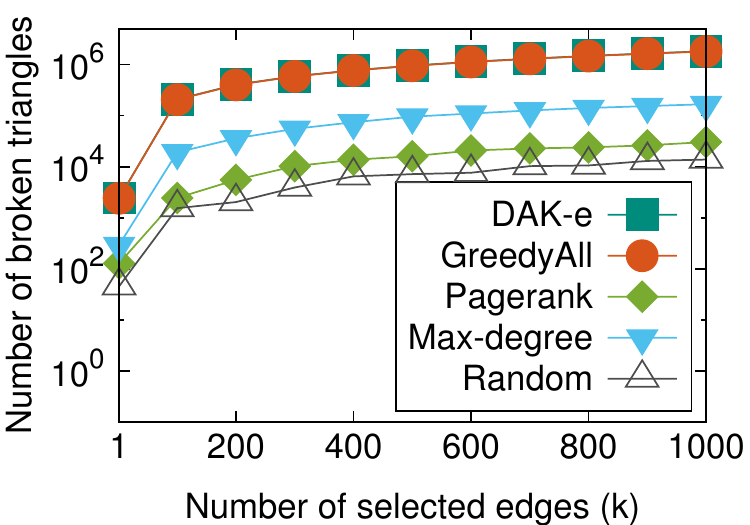}}
	\subfloat[Gnutella]{\includegraphics[width=0.33\textwidth]{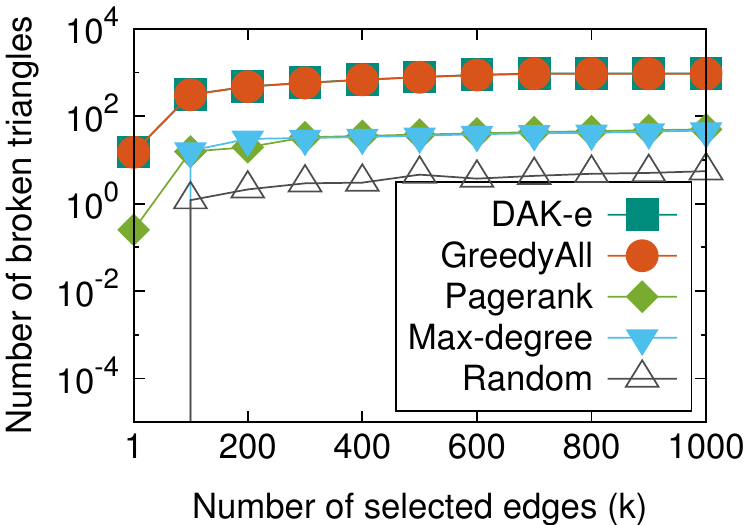}}\subfloat[Google]{\includegraphics[width=0.33\textwidth]{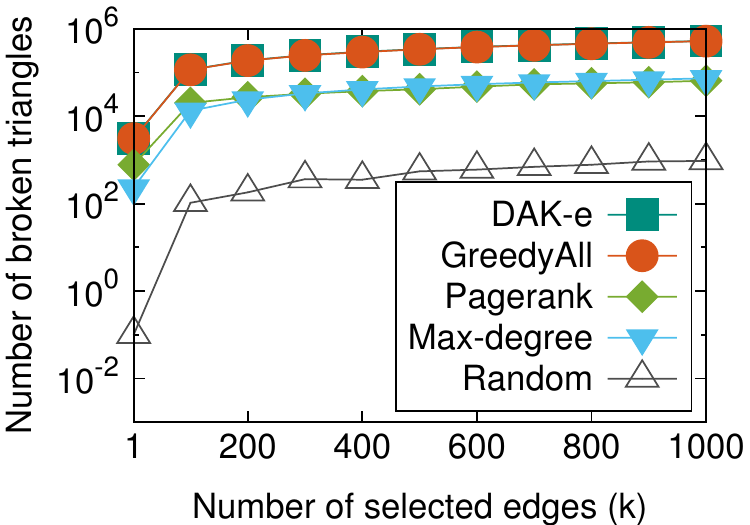}}\vspace{-0.1in}\
	\subfloat[Skitter]{\includegraphics[width=0.33\textwidth]{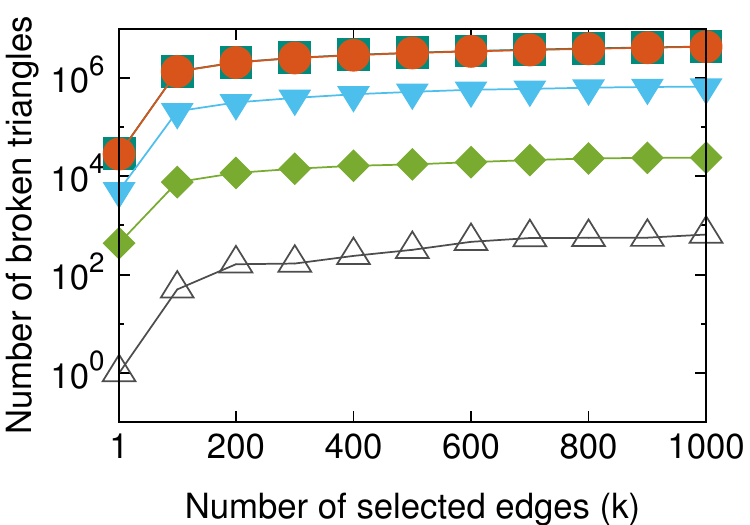}}
	\subfloat[Wiki-Talk]{\includegraphics[width=0.33\textwidth]{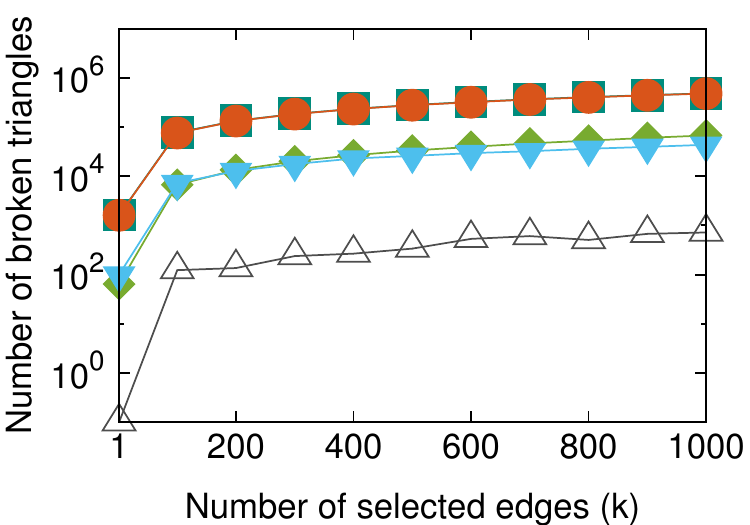}}
	\subfloat[Orkut]{\includegraphics[width=0.33\textwidth]{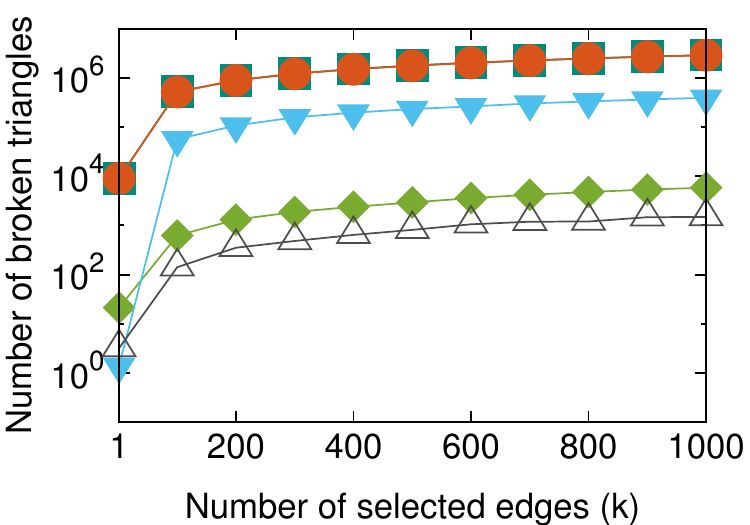}}
	\caption{Number of broken triangles broken by edge removals (higher value is better)}
	\label{fig:edge_tri}
\end{figure*}

\begin{figure*}[ht!]
	\centering
	\subfloat[Flickr]{\includegraphics[width=0.33\textwidth]{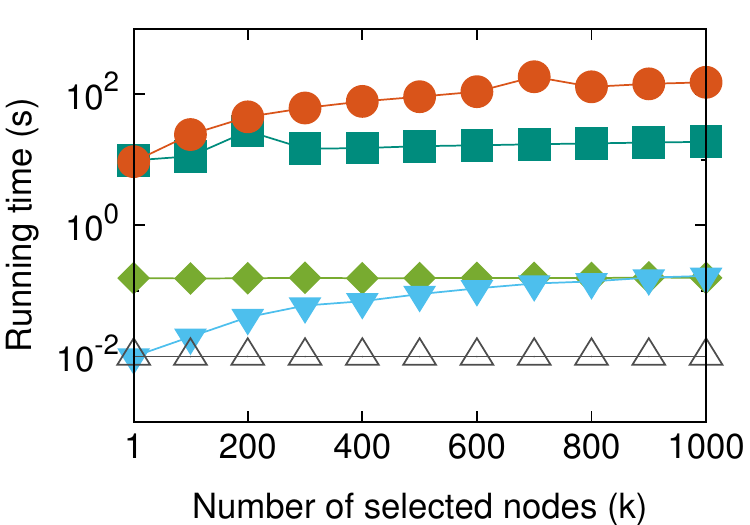}}
	\subfloat[Gnutella]{\includegraphics[width=0.33\textwidth]{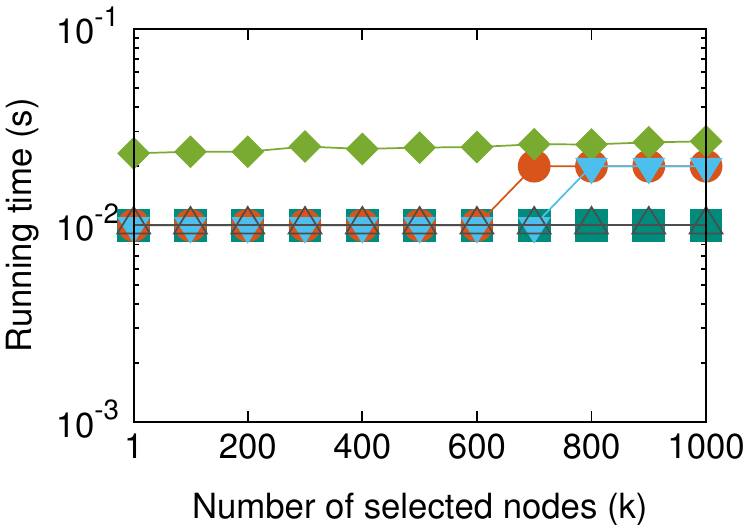}}\subfloat[Google]{\includegraphics[width=0.33\textwidth]{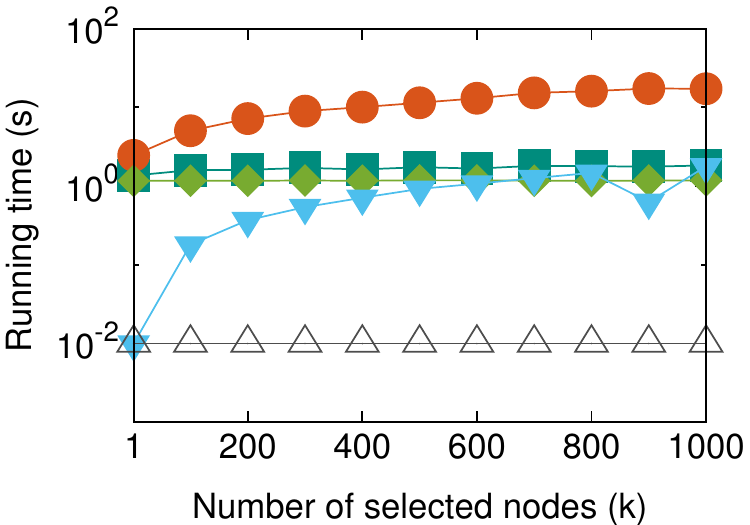}}\vspace{-0.1in}\
	\subfloat[Skitter]{\includegraphics[width=0.33\textwidth]{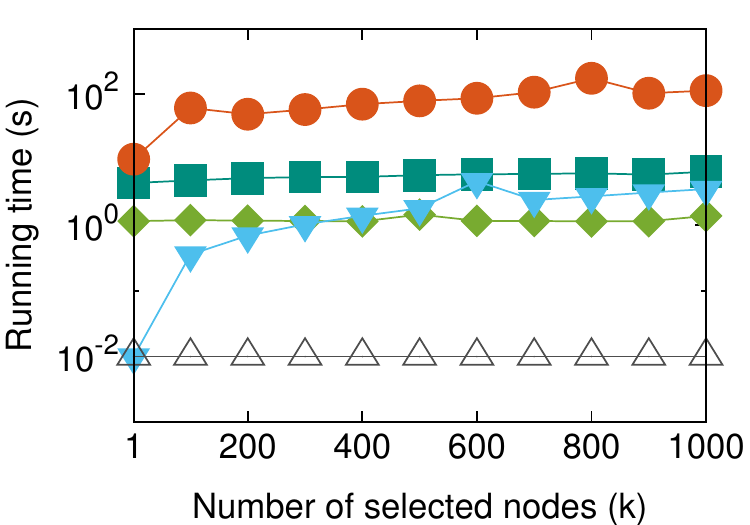}}
	\subfloat[Wiki-Talk]{\includegraphics[width=0.33\textwidth]{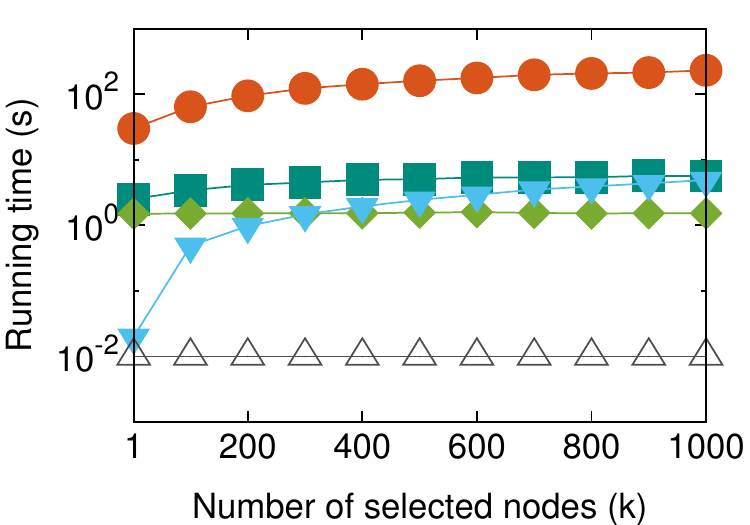}}
	\subfloat[Orkut]{\includegraphics[width=0.33\textwidth]{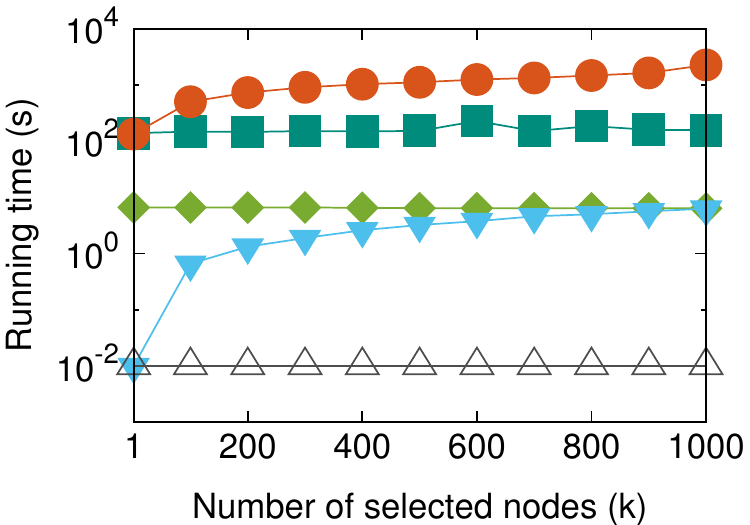}}
	\caption{Running time of node removal algorithms (legends in Fig.~\ref{fig:node_tri})}
	\label{fig:node_time}
\end{figure*}

\begin{figure*}[ht!]
	\centering
	\subfloat[Flickr]{\includegraphics[width=0.33\textwidth]{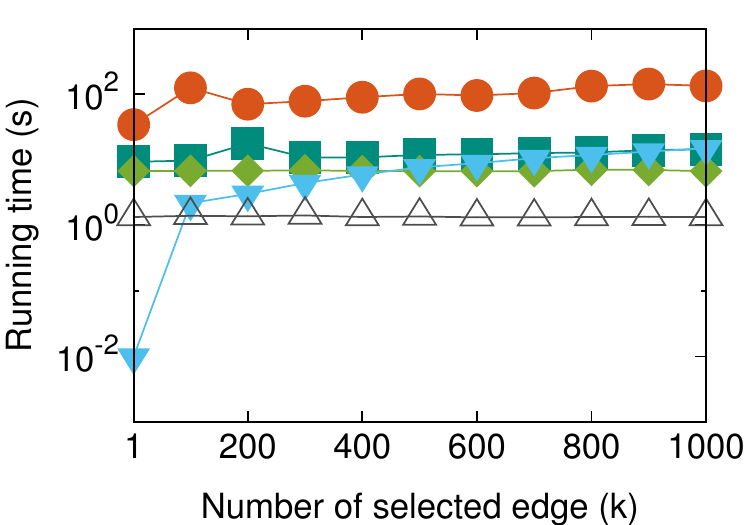}}
	\subfloat[Gnutella]{\includegraphics[width=0.33\textwidth]{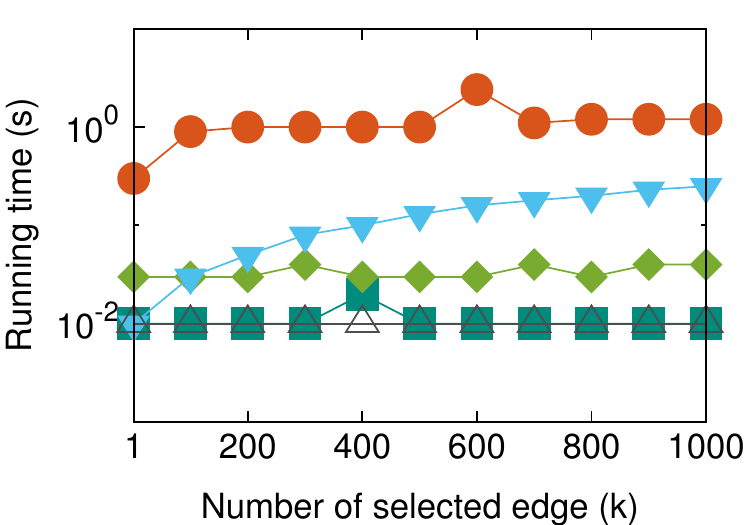}}\subfloat[Google]{\includegraphics[width=0.33\textwidth]{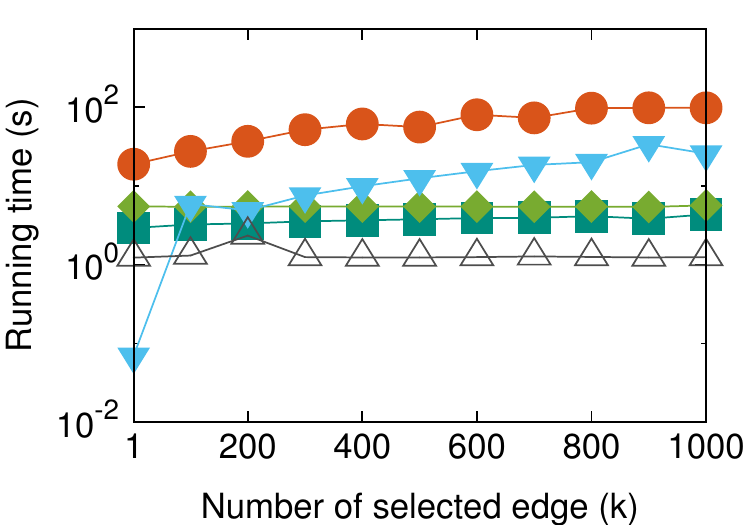}}\vspace{-0.1in}\
	\subfloat[Skitter]{\includegraphics[width=0.33\textwidth]{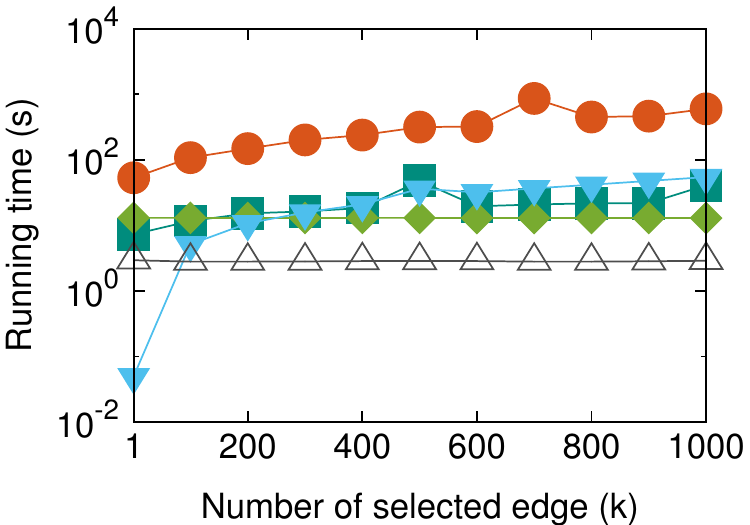}}
	\subfloat[Wiki-Talk]{\includegraphics[width=0.33\textwidth]{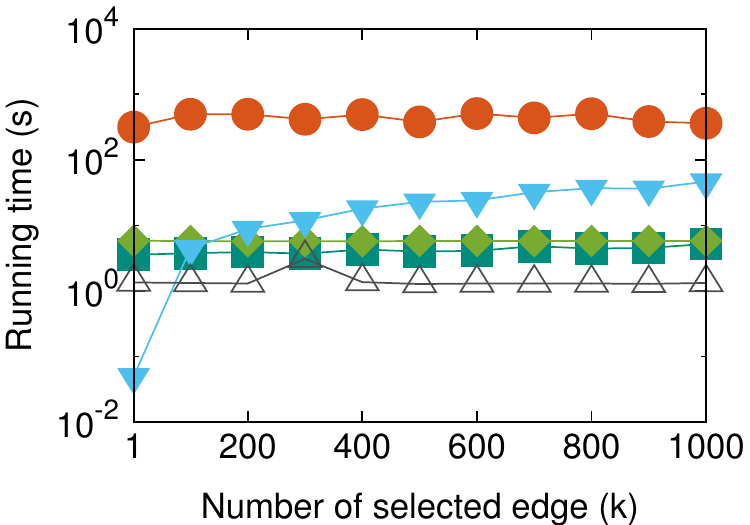}}
	\subfloat[Orkut]{\includegraphics[width=0.33\textwidth]{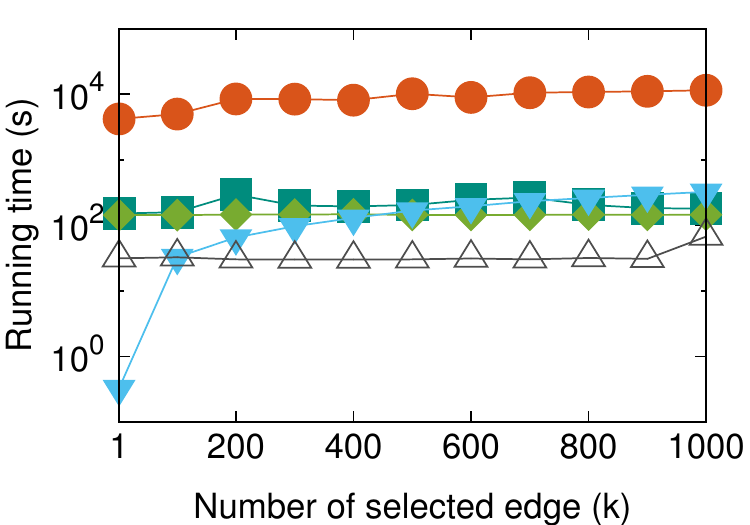}}
	\caption{Running time of edge removal algorithms (legends in Fig.~\ref{fig:node_tri})}
	\label{fig:edge_time}
\end{figure*}